\documentclass[aps,amssymb,prl,twocolumn,showpacs]{revtex4}
\usepackage{epsfig}
\usepackage{array}

\def\al{{\it et al}}

\def\0{\varnothing}

\def\Tr{{\rm Tr}}

\begin{document}
\title{Criterion for phase separation in one-dimensional driven systems}
\author{Y. Kafri$^{(a)}$,E. Levine$^{(a)}$,D. Mukamel$^{(a)}$,G. M. Sch\"{u}tz$^{(b)}$,J. T\"or\"ok$^{(a)}$}
\affiliation{$(a)$ Department of physics of complex systems,
Weizmann Institute of Science, Rehovot, Israel 76100.\\
$(b)$Institut f\"{u}r Festk\"orperforschung, Forschungszentrum
J\"{u}lich, 52425 J\"{u}lich, Germany.}

\date{April 15, 2002}
\begin{abstract}
A general criterion for the existence of phase separation in
driven density-conserving one-dimensional systems is proposed. It
is suggested that phase separation is related to the size
dependence of the steady-state currents of domains in the system.
A quantitative criterion for the existence of phase separation is
conjectured using a correspondence made between driven diffusive
models and zero-range processes. The criterion is verified in all
cases where analytical results are available, and predictions for
other models are provided.
\end{abstract}

\maketitle

The existence of phase separation and spontaneous symmetry
breaking in low-dimensional systems far from thermal equilibrium
has been a subject of recent interest~\cite{Evans95,Mukamel00}.
While it is well known that these phenomena do not take place in
one dimension in thermal equilibrium, several models of driven one
dimensional systems with local dynamics have recently been
demonstrated to exhibit both~\cite{Evans98,Rittenberg99,LBR00}.
Whether or not a given model exhibits phase separation is in many
cases not a simple question to answer, and it may depend on
numerical evidence which could be rather subtle.

For example, in a recent 3-species model introduced by Arndt
\al~\cite{Rittenberg99} (AHR), it has been suggested that one
should expect two distinct phase separated states: one in which
the three species are fully separated from each other (related to
the phase separation observed by Evans \al~\cite{Evans98} in a
related model), and the other is a more subtle mixed state whose
existence is supported by extensive numerical simulations of
systems of finite length and by a mean-field treatment.
Subsequently, an analytical analysis of the model has shown that
the mixed state is in fact disordered, and that in order to see
this in simulations one has to study extremely long systems (of
the order of $10^{70}$), far beyond existing numerical
capabilities~\cite{Sasamoto00}.

In another example introduced by Korniss \al~a two lane extension
of a 3-species driven system was studied~\cite{Korniss}. It has
been suggested that while for this model the one lane system does
not exhibit phase separation~\cite{Sandow}, this phenomenon does
exist in the two lane model. The studies rely on numerical
simulations of systems of length up to $10^4$. This result is
rather surprising and not well understood. It may very well be the
case that as for the AHR model, the two lane model does not
actually exhibit phase separation in the thermodynamic limit, and
that this could be seen only by studying extremely long systems.
It would thus be of great importance to find other criteria, which
could distinguish between models supporting phase separation from
those which do not.

In this Letter we introduce a simple general criterion for the
existence of phase separation in density-conserving
one-dimensional driven systems. Phase separation is usually
accompanied by a coarsening process in which small domains of,
say, the high density phase coalesce, eventually leading to
macroscopic phase separation. This process takes place as domains
exchange particles through their currents. When smaller domains
exchange particles with the environment with faster rates than
larger domains, a coarsening process is expected, which may lead
to phase separation. Our criterion quantifies this mechanism, and
relates the existence of phase separation to the steady-state
currents through which domains exchange particles. The criterion
is readily applicable even in cases which cannot be decided by
direct numerical simulations.

In order to explicitly state the criterion we note that in many
models which carry a non-zero current in the thermodynamic limit
the current of a finite domain of size $n$ takes the form $J_n =
J_\infty (1+b/n)$ to leading order in $1/n$. For $b>0$ the current
of long domains is smaller than that of short ones, which leads to
a tendency of the longer domains to grow at the expense of smaller
ones. According to our criterion phase separation takes place at
high-densities only for $b>2$. Moreover if the current decays to
its asymptotic value as $J_n = J_\infty (1+b/n^{\sigma})$, the
model is predicted to phase separate at any density for $\sigma<1$
while it is always homogeneous for $\sigma>1$. In some models
$J_\infty =0$, although, due to lack of detailed balance, the
current $J_n$ of a finite system is non-vanishing. In this case
the system is predicted to phase separate at any density. Models
for which $J_n$ decays exponentially to zero with $n$ have been
analyzed in the past and indeed were shown to exhibit phase
separation~\cite{Evans98,Rittenberg99,LBR00}.

The results presented above emerge from a careful analysis of a
zero-range process (ZRP) which could be viewed as a generic model
for domain dynamics in one-dimension. To define this process we
consider a one-dimensional lattice of $M$ sites, or ``boxes'',
with periodic boundary conditions. Particles, or ``balls'', are
distributed among the boxes with the box $i$ occupied by $n_i$
balls. The dynamics is defined in the following way: a box $i$ is
chosen at random and a particle is removed from it and transferred
to a left (right) neighbor with rates $p w_{n_i}$ ($(1-p)
w_{n_i}$) where $0 \leq p \leq 1$. The rate $w_{n_i}$ depends only
on the number of balls in that box. The model may either be
unbiased ($p=1/2$), or biased ($p \neq 1/2$).

In a grand canonical ensemble, namely an ensemble where the number
of boxes $M$ is fixed while the number of balls is allowed to
fluctuate with their average number controlled by a fugacity $z$,
the steady state weight of a configuration of the ZRP is known to
be~\cite{Schutz01}
\begin{equation}
\label{eq:wbar} W_{\mbox{\footnotesize ZRP}} \left( \lbrace n_i
\rbrace \right) = \prod_{i=1}^{M}{z^{n_i}{\cal F}_{n_i}}\;.
\end{equation}
Here ${\cal F}_k = \prod_{m=1}^{k}{1/w_{m}}$ for $k \geq 1$ and
${\cal F}_0 =1$. In this ensemble boxes are statistically
independent with a single-site occupation distribution function
given by $P(k) \sim z^k {\cal F}_k$. Depending on the rates $w_n$
the model may or may not exhibit condensation in the thermodynamic
limit, whereby the occupation number of one of the boxes becomes
macroscopically large. Clearly the rate $w_n$ must be a decreasing
function of $n$ in order for larger blocks to be favored and to
support condensation. It is known~\cite{Schutz01} that
condensation occurs at any density when $w_n \to 0$ with $n \to
\infty$, or when it decreases to a non-vanishing asymptotic value
$w_\infty$ as $w_\infty\left(1+b/n^\sigma\right)$ with $\sigma<1$;
no phase separation takes place for $\sigma>1$; for $\sigma=1$
phase separation takes place at high densities only for $b>2$.

This model may be used to gain physical insight into the dynamics
of driven one-dimensional systems. Occupied boxes represent
domains of the high density phase. The currents leaving domains
are represented by the rates of the ZRP. This is done by
identifying the rate $w_n$ associated with a box containing $n$
balls with the currents $J_n$ leaving a domain of $n$ particles. A
bias in the currents to a certain direction may be incorporated
through $p$ as defined above. The existence of a box with a
macroscopic occupation in the ZRP corresponds to phase separation
in the driven model.

In the following we consider several one-dimensional driven
systems and study their domain dynamics by introducing a
corresponding zero-range process. By analyzing the ZRP, the
existence of phase separation in the original model may be
addressed. We begin by considering the AHR model. We show that for
this model the corresponding ZRP yields its {\it exact}
steady-state domain-size distribution. This ZRP does not exhibit
phase separation, in agreement with the results of Rajewsky
\al~\cite{Sasamoto00}. We then discuss the two-lane
model~\cite{Korniss} and argue that it, too, does not exhibit
phase separation contrary to results of numerical simulations of
finite systems.

The AHR model is a three-state model on a ring. Each site is
either empty ($0$), occupied by a positive ($+$) or a negative
($-$) particle. The model evolves by a random sequential dynamics
in which a pair of nearest neighbor sites is chosen at random and
exchanged with the rates:
\begin{equation}
\label{eq:rates} +\;0 \mathop{\rightarrow}\limits^{\alpha}  0\;+
\;\;\;;\;\;\; 0\,- \mathop{\rightarrow}\limits^{\alpha}  -\,0
\;\;\;;\;\;\; +\,- \mathop{\rightleftarrows}\limits^{1}_{q} -\,+
\;.
\end{equation}
This dynamics conserves the number of particles of each type. As
in most studies of this model we consider equal densities of
positive and negative particles. Numerical simulations suggest
that the model has three states~\cite{Rittenberg99}: a fully
ordered state for $q>1$, in which the system strongly phase
separates into its three phases; a mixed state for $q_c<q<1$ in
which the particles (both positive and negative) in the system
condense into a single high density phase separated from a low
density gas-like phase; and a disordered state for $q<q_c$ where
particles and vacancies are homogeneously distributed. On the
other hand exact calculations within the grand canonical
ensemble~\cite{Sasamoto00} show that the mixed state is in fact
disordered, with a finite average length of the high density
domains in the thermodynamic limit. Thus the homogeneous and the
mixed states constitute a single disordered state. The system
therefore exhibits only two states, one fully phase separated for
$q>1$ and the other disordered for $q<1$.

It is known~\cite{Rittenberg99,Sasamoto00} that for this model the
steady-state weight, $W_L$(${\cal C}$), of a given microscopic
configuration ${\cal C}$ is
\begin{equation}
W_L \left( {\cal C} \right) = \Tr \prod_{i=1}^{L} \left[ z \left(
\delta_{\tau_i (+)} D + \delta_{\tau_i (-)} E \right) +
\delta_{\tau_i (0)} A \right]\;.
\end{equation}
Here $L$ is the length of the lattice, $\tau_j=+,-,0$ when site
$j$ is occupied by a $+,-$ or $0$, respectively and $z$ is the
fugacity which controls the average number of particles. The
fugacity is the same for positive and negative particles so their
average number is equal. The matrices $D, E$ and $A$ satisfy
\begin{equation}
\label{eq:algebra} DE-qED = D+E \quad \quad \alpha DA = \alpha AE
= A
\end{equation}
to give the correct steady-state weight. Explicit representations
which satisfy this algebra are
known~\cite{Rittenberg99,Sasamoto00}. For our purpose it is
sufficient to note that there exists a representation in which
$A=\left|0\right>\left<0\right|$ is a projection operator
satisfying $A^2=A$, where $\left|0\right>$ is a state vector.

\begin{figure}
\centerline{\epsfig{file=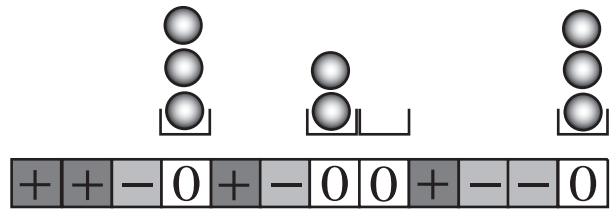,width=8truecm}}
\caption{\label{fig:cartoon} A typical configuration of the
three-state model (bottom) and its corresponding configuration in
the ZRP (top). Periodic boundary conditions are imposed on the two
models. }
\end{figure}

We now consider the steady state of this model in an ensemble in
which the number of vacancies $M$ is held fixed while the number
of particles, and thus also the length $L$ of the lattice are
allowed to fluctuate. We refer to this ensemble as a grand
canonical. A typical configuration of the model consists of blocks
of particles bounded by vacancies. By a block we mean an
uninterrupted sequence of positive and negative particles, bounded
between two zeros. Let $n_i$~($i=1, \ldots, M$) be the length of
the $i$th block located to the left of the $i$th vacancy. The
block lengths can take the values $n_i=0, \ldots ,L-M$ and satisfy
$\sum_{i} n_i = L-M$. The partial trace $W( \lbrace n_i \rbrace)$
of all weights of microscopic configurations $W_L(\cal C)$
consistent with $n_1, \ldots, n_M$, takes the form
\begin{equation}
W( \lbrace n_i \rbrace)=\prod_{i=1}^{M}{z^{n_i}{\cal Z}_{n_i}}\;,
\label{eq:mweight}
\end{equation}
where ${\cal Z}_k =\left<0\right|(D+E)^{k}\left|0\right>$ is the
sum over all weights of microscopic configurations of a block of
length $k$. Here we have used the above representation of the
matrix $A$. Within the grand canonical ensemble the various
domains are statistically independent with a domain size
distribution $P(k) \sim z^k {\cal Z}_k$. It is known that ${\cal
Z}_k$, with the algebra (\ref{eq:algebra}), is the partition
function (sum over all weights) of the partially asymmetric
exclusion process (PASEP) on a one-dimensional lattice of $k$
sites and open boundary conditions \cite{Sasamoto99,Blythe00}. The
boundary conditions are such that the positive particles are
injected at rate $\alpha$ at the left end and are removed from the
right end with the same rate. The dynamics in the bulk of the
system is given by the same rates as in (\ref{eq:rates}).
Moreover, the current in such a system of size $k$ is given by
$J_k = {\cal Z}_{k-1}/{\cal Z}_k$ so that ${\cal
Z}_k=\prod_{m=1}^{k} 1/J_m$ for $k \geq 1$.

We now turn to define the corresponding ZRP. We consider a ZRP in
which each box represents a vacancy in the AHR model. The box $i$
is occupied by $n_i$ ``balls'', which corresponds to the number of
particles (positive and negative) in the block to the left of the
$i$th vacancy (see Fig. \ref{fig:cartoon}). Since in the AHR model
the positive and negative currents of a block of size $n$ are
equal to $J_n$, we choose the hopping rates of the ZRP to be $w_n
= 2 J_n$ with a symmetric hopping to the right and to the left.
Here again $J_n$ is the current of the PASEP in an open system of
size $n$. Thus for this ZRP one has ${\cal F}_k=2^{-k}{\cal Z}_k$.
Hence the weight of a given block configuration $n_1, \ldots, n_M$
in the AHR and the corresponding configuration of the ZRP are the
same (up to a $2^{-k}$ prefactor which can be absorbed into the
definition of the fugacity $z$)~\cite{com}. The ZRP provides a
simple interpretation of the AHR steady-state dynamics in which
blocks interact via exchange of particles. Each block in the AHR
behaves as a PASEP with open boundary conditions and equal
injection and ejection rates at steady state. Neighboring blocks
exchange particles at a rate given by the PASEP current.

We now use the ZRP to study the block size distribution in the AHR
model. To do this we use the asymptotic form of the current of the
PASEP with $q<1$,
\begin{equation}
J_n = J_\infty\left(1+b/n+{\cal O}\left(1/n^2\right)\right)\;,
\label{eq:current}
\end{equation}
with $J_\infty = (1-q)/4$, $b=3/2$ for $1>q>1-2\alpha$ and $b=-1$
for $q<1-2 \alpha$~\cite{Sasamoto99,Blythe00}. Therefore, since
$b<2$ in both regimes, no condensate can appear for $q<1$, in
agreement with ~\cite{Sasamoto00}. Moreover, using the form
(\ref{eq:current}) of $J_n$ it is easy to show that the block size
distribution in the homogeneous phase is given by
\begin{equation}
\label{eq:dist} P(k) \sim
\frac{1}{k^b}\exp(-k/\xi)\;\;;\;\;\xi=\frac{1}{|\ln
(z/J_\infty)|}\;.
\end{equation}
It is evident that for $b \leq 2$ the average block size diverges
as $\xi \to \infty$ and the distribution function is valid for any
density no matter how large. No phase transition takes place in
this case. However, for $b>2$ the average block size remains
finite for $\xi \to \infty$ necessitating the existence of a phase
transition, which results in a macroscopic block at high
densities. Note that at $q=1-2\alpha$ the block-size distribution
function changes in a non-analytic manner. This point was first
noted for the case of $q=0$ in~\cite{Sandow}.

This result yields interesting insight into the origin of the
apparent transition seen in simulations whereby in a certain $q$
interval the correlation length $\xi$ becomes exceedingly large.
It can be shown~\cite{TBP} that the $q$ dependence of the
correlation length $\xi$ is introduced by the higher order
corrections (for example $c/n^2$) to the
current~(\ref{eq:current}). We have calculated the correlation
length $\xi$ of a ZRP with rates $w_n=1+3/2n+c/n^2$ for a given
density as a function of the parameter $c$ (corresponding to
changing $q$)~\cite{TBP}. We find that $\xi$ exhibits a sharp
increase of a few orders of magnitude over a narrow range of
values of $c$. This reflects itself in large (but finite) blocks
and an apparent phase separation in direct simulations.

The physical picture emerging from this analysis offers a rather
robust mechanism for phase separation, and we conjecture that it
has a more general validity. We expect it to apply to other
conserving driven models even though an exact correspondence to
the ZRP may not be evident.

\begin{figure}
\centerline{\epsfig{file=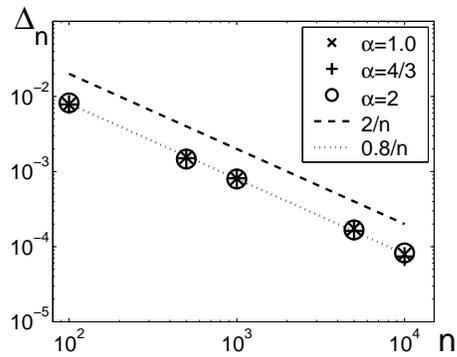,width=6truecm}}
\caption{\label{fig:twolane} The finite-size corrections to the
current $\Delta_n=(J_n-J_\infty)/J_\infty$ in the two-lane model
with open boundary conditions, for different system sizes. Here
$\gamma=1$.}
\end{figure}

We now demonstrate the use of the conjecture for the two-lane
model introduced by Korniss \al~\cite{Korniss}. This model is a
generalization of (\ref{eq:rates}) with $q=0$ to two lanes. Here
in addition to the hopping process (\ref{eq:rates}) within each
lane, particles may hop to neighboring empty sites on the other
lane with rate $\gamma \alpha$, and to exchange with a neighboring
particle on the other lane with rate $\gamma$. Numerical studies
of the model have suggested that for large enough $\alpha$ the
system phase separates~\cite{Korniss}. In this state the particles
condense into a single high density block. However, physical
insight into the phenomenon is lacking. In particular it is not
understood why the two-lane model seems to exhibit phase
separation while its singled-lane version does not~\cite{Sandow}.
In order to apply the conjecture, the current $J_n$ of a block of
size $n$ is calculated for the two lane model. This is done by
considering an open two lane system of length $n$ with no
vacancies where particles are injected and ejected at the
boundaries with equal rates $\alpha$. Setting $\gamma=1$ as
in~\cite{Korniss} we show the results of numerical simulations for
several values of $\alpha$ in Fig. \ref{fig:twolane}. By comparing
the results for the corrections to the current
$\Delta_n=(J_n-J_\infty)/J_\infty$ with the line $2/n$ it is easy
to see that in this case $b<2$. In fact, the curve may be best
fitted to $b \simeq 0.8$. Our conjecture thus implies that phase
separation does not take place in this model. Note that even if
one tries to fit $\Delta_n$ to $1/n^\sigma$ with $\sigma \neq 1$
one finds that $\sigma \gtrsim 1$ yielding the same conclusion. We
note that simulating the open systems and obtaining the asymptotic
behavior of the current involves a relatively modest numerical
effort, as one only needs to simulate rather small systems. This
should be compared with the huge systems which are needed in order
to demonstrate the lack of phase separation in direct simulations.
Indeed, the conjecture suggests that the apparent phase separation
found in numerical studies is due to simulations of systems much
smaller than the typical domain size.

We have studied another extension of model (\ref{eq:rates}),
whereby the hopping rates depend on both nearest and next-nearest
neighbors. In these models we have shown that $b = 3/2$ with
$\sigma = 1$. The conjecture implies no phase separation, in
contrast to our direct simulations.

We now apply the conjecture to a class of models with vanishing
$J_\infty$. To this end, consider model (\ref{eq:rates}) with
$q>1$. As before, we define a block as a sequence of particles,
both positive and negative, bounded by two vacancies. The
corresponding open system is the PASEP with $q>1$, with open
boundary conditions and equal injection and ejection rates,
studied in~\cite{Blythe00}. The current in such a block, as in the
corresponding open system, is exponentially decreasing with the
block size, $J_n \sim q^{-n/2}$. This is easily understood, as the
particles are moving against the bias when $q>1$. Using the
conjecture the system is expected to phase separate for any
density. Several similar models with exponentially decaying
currents have been shown to exhibit a strongly phase separated
state~\cite{Evans98,Rittenberg99,LBR00}.

We end by reiterating the assumptions involved in the conjecture,
which relates phase separation to the currents $J_n$ of finite
blocks. It is assumed that the current flowing through a block is
given by its steady-state value and is independent of its
neighboring blocks. This may be justified by the fact that the
coarsening time of large domains is very long, and the domains
have a chance to equilibrate long before they coarsen.

Although the criterion introduced in this Letter has not been
proved to hold in general, its underlying physical mechanism for
phase separation is rather robust, suggesting a broad
applicability. It is of interest to test the conjecture on other
models to check its general validity. In the models exhibiting
phase separation we have studied so far, the current $J_\infty$
vanishes. It would be interesting to find a one-dimensional driven
model for which $J_\infty > 0$ and $b>2$. According to the
conjecture such a model should exhibit a novel type of phase
separation.

\begin{acknowledgments} We thank M. R. Evans for many discussions and
suggestions and C. Godr\`eche for very fruitful discussions on the
AHR model. He and S. Sandow are thanked for communicating to us a
preliminary version of \cite{Sandow}. We also thank I. Kanter and
V. Rittenberg for discussions. The support of the Israeli Science
Foundation is gratefully acknowledged. GMS thanks the Einstein
center and Deutsche Forschungsgemeinschaft for support.
\end{acknowledgments}


\begin{thebibliography}{999}

\bibitem{Evans95} M. R. Evans, D. P. Foster, C. Godr\`eche, and D.
Mukamel, Phys. Rev. Lett. {\bf 74}, 208 (1995).

\bibitem{Mukamel00} For a recent review see D. Mukamel in {\it Soft and Fragile Matter:
Nonequilibrium Dynamics, Metastability and Flow}, Eds. M.E. Cates
and M.R. Evans (Institute of Physics Publishing, Bristol, 2000);
{\tt cond-mat/0003424}.

\bibitem{Evans98} M. R. Evans, Y. Kafri, H. M. Koduvely, and D. Mukamel,
Phys. Rev. Lett. {\bf 80}, 425 (1998); Phys. Rev. E {\bf 58}, 2764
(1998).

\bibitem{Rittenberg99} P. F. Arndt, T. Heinzel, V. Rittenberg, J. Phys. A
{\bf 31}, L45 (1998); J. Stat. Phys. {\bf 97}, 1 (1999).

\bibitem{LBR00} R. Lahiri, S. Ramaswamy, Phys. Rev. Lett. {\bf 79}, 1150 (1997);
R. Lahiri, M. Barma, S. Ramaswamy, Phys. Rev. E {\bf 61}, 1648
(2000).

\bibitem{Sasamoto00} N. Rajewsky, T. Sasamoto, E. R. Speer, Physica A
{\bf 279}, 123 (2000); T. Sasamoto, and D. Zagier, J. Phys. A {\bf
34}, 5033 (2001).

\bibitem{Korniss} G. Korniss, B. Schmittmann and R. K. P. Zia,
Europhys. Lett. {\bf 45}, 431 (1999); J. T. Mettetal, B.
Schmittman and R. K. P. Zia, {\tt cond-mat/0110301}.

\bibitem{Sandow} C. Godr\`eche and S. Sandow , unpublished.

\bibitem{Schutz01} See, {\it e.g.}, F. Spitzer, Advances in Math. {\bf 5}, 246
(1970); M. R. Evans, Braz. J. Phys. {\bf 30}, 42 (2000); O. J.
O'Loan, M. R. Evans and M. E. Cates, Phys. Rev. E {\bf 58}, 1404
(1998).

\bibitem{Sasamoto99} T. Sasamoto, J. Phys. A {\bf 32}, 7109 (1999).

\bibitem{Blythe00} R. A. Blythe, M. R. Evans, F. Colaiori, and F. H. L. Essler, J.
Phys. A {\bf 33}, 2313 (2000).

\bibitem{com} Note that each block configuration in the AHR model
carries an extra degeneracy which is bounded from above by $L$ as
compared to the corresponding ZRP. This degeneracy corresponds to
the number of ways of placing a given block configuration on a
lattice of length $L$. It does not affect the results in the
thermodynamic limit.

\bibitem{TBP} Y. Kafri, E. Levine, D. Mukamel and J. T\"or\"ok, J. Phys. A in press; {\tt
cond-mat/0206145}.


\end{thebibliography}
\end{document}